\documentclass[aps,preprint,showpacs,floats,epsf,epsfig,nofootinbib,12pt]{revtex4}
\usepackage{graphicx}
\usepackage{epsfig}
\usepackage[dvips]{color}

\def\be{\begin{eqnarray}}
\def\en{\end{eqnarray}}
\def\non{\nonumber}
\def\la{\langle}
\def\ra{\rangle}
\def\pp{{\prime\prime}}

\def\vp{\varepsilon}

\begin{document}

\title{Determining the effective Wilson coefficient $a_2$ in terms of $BR(B_s\to \eta_c\phi)$ and evaluating $BR(B_s\to \eta_cf_0(980))$}

\vspace{1cm}

\author{ Hong-Wei Ke$^{1}$\footnote{khw020056@hotmail.com} and
        Xue-Qian Li$^2$\footnote{lixq@nankai.edu.cn}  }

\affiliation{  $^{1}$ School of Science, Tianjin University,
Tianjin 300072, China
\\
  $^{2}$ School of Physics, Nankai University, Tianjin 300071, China
   }

\vspace{12cm}

\begin{abstract}
In this work, we investigate decays of $B_s\to \eta_c\phi$ and
$B_s\to \eta_cf_0(980)$ in a theoretical framework. The
calculation is based on the postulation that $f_0(980)$ and
$f_0(500)$ are mixtures of pure quark states ${1\over\sqrt 2}
(\bar uu+\bar dd)$ and $\bar ss$. The hadronic matrix elements for
$B_s\to \phi$ and $B_s\to f_0(980)$ are calculated in the
light-front quark model and the important Wilson coefficient $a_2$
which is closely related to non-perturbative QCD is extracted.
However, our numerical results indicate that no matter how to
adjust the mixing parameter to reconcile contributions of
$f_0(980)$ and $f_0(500)$, one cannot make the theoretical
prediction on $B_s\to \eta_c+\pi^+\pi^-$ to meet the data.
Moreover, the new measurement of $BR(B_s\to
J/\psi+f_0(500))<1.7\times 10^{-6}$ also negates the mixture
scenario. Thus, we conclude that the recent data suggest that
$f_0(980)$ is a four quark state ( tetraquark or $K\bar K$
molecule ), at least the fraction of its pure quark constituents
is small.

\pacs{13.25.Hw, 14.40.Cs, 12.39.Ki}

\end{abstract}

\maketitle

\section{Introduction}
The values of $BR(B_s\to \eta_c\phi)=(5.01 \pm 0.53 \pm 0.27
\pm0.63)\times 10^{-4}$ and $BR(B_s\to \eta_c\pi^+\pi^-)=1.76 \pm
0.59 \pm 0.12 \pm 0.29)\times 10^{-4}$ recently measured by the
LHCb collaboration \cite{Aaij:2017hfc} have stimulated new vigor
for studying the hadron structures and the decay mechanism which
is closely related to the non-perturbative QCD
effects. Based on data, the Collaboration suggests that the
$\pi^+\pi^-$ pair in $B_s\to \eta_c\pi^+\pi^-$ arises from the
decay of $f_0(980)$. To understand the data and look for some
hints about involved physics, corresponding theoretical
calculations are needed. The traditional scheme is using the heavy quark
effective theory (HQET) \cite{Isgur:1989vq,Georgi:1990um} and naive factorization which is
an old issue, but still applicable in parallel to the fancy
theories such as SCET and others.

The subprocess is $b\to c\bar c s$,
and at the tree level, the main contribution is the internal $W-$emission
while the light quark serves as a spectator. For a completeness,
let us briefly retrospect the standard procedures of applying HQET.
In the
HQET, the corresponding lagrangian is written as
\begin{equation}
{\cal L}=c_1\bar c\gamma_{\mu}(1-\gamma_5)b\bar s\gamma^{\mu}(1-\gamma_5)c+c_2
\bar c\gamma_{\mu}(1-\gamma_5)c\bar s\gamma^{\mu}(1-\gamma_5)b,
\end{equation}
where $c_1={1\over 2}(c_++c_-)$ and $c_2={1\over 2}(c_+-c_-)$ and
$c_+,c_-$ are obtained by means of the re-normalization group
equation (RGE). Sandwiching the lagrangian between the initial and
final states, we have
\begin{eqnarray}\label{factor}
&& <\eta_c \phi(f_0(980)|{\cal L}|B_s> \nonumber\\
&=& a_2\left(<\eta_c\phi (f_0(980))| \bar
c\gamma_{\mu}(1-\gamma_5)c|0><0|\bar
s\gamma^{\mu}(1-\gamma_5)b|B_s>\right.\nonumber\\&& \left.
+<\eta_c|\bar c\gamma_{\mu}(1-\gamma_5)c|0><\phi(f_0(980))|\bar
s\gamma^{\mu}(1-\gamma_5)b|B_s>\right).
\end{eqnarray}

It is noted that the $c_1O_1$ term contributes to the decay
process via a color-re-arrangement. Naively, one can expect
$a_2=c_2+1/3 c_1$ by the color rearrangement. However, it was
pointed out by some authors
\cite{Bauer:1986bm,Cheng:1986an,Li:1988hr}``the sub-leading order
in $1/N_c$ includes not only the next-to-leading vacuum-insertion
contribution but also the nonperturbative QCD correction" .
Keeping the factorization form, one should replace $a_2=c_2+
c_1/3$ by $a_2=c_2+ c_1/3+\epsilon_a/2$ where $\epsilon_a$ is a
parameter(with Cheng's notation\cite{Cheng:1986an}). Even though
one can calculate $\epsilon_a$ in terms of some models
\cite{Li:1988hr}, the result is not accurate, therefore, generally
one should phenomenologically fix it by fitting the well measured
data. Our work is exactly along the line. This issue was first
discussed in Ref.\cite{Bauer:1986bm}. In fact, $a_2$ includes some
non-perturbative QCD effects so it is not universal for the
different channels of the D or B decays\cite{Cheng:1994fr} as
shown above. Definitely, determining the value of $a_2$ based on
data fitting one can obtain information about non-perturbative
physics. In Ref.\cite{Cheng:1994fr} $a_2=0.23\pm0.06$ was fixed by
fitting $BR(B\to D^{(*)}\pi(\rho))$ . In this work we instead use
$B_s\to \eta_c\phi$ to extract the corresponding $a_2$ value. Then
we evaluate $BR(B_s\to \eta_c f_0(980))$ in terms of the newly
obtained $a_2$. It is worth of noticing that the derivation is
based on the postulation that $f_0(980)$ is of a pure $\bar qq$
structure ($q$ stands as u,d and s quarks). We will come to this
issue  for some details in the last section.

In order to calculate the decay width under the factorization assumption
one needs to evaluate the hadronic transition matrix element between two
mesons. Since the transition is governed by the non-perturbative QCD effects,
so far one has to invoke certain phenomenological models.
In this work, we employ the light-front quark model(LFQM).
This relativistic model
has been thoroughly discussed in literatures
\cite{Jaus:1999zv,Cheng:2003sm} and applied to study several hadronic
transition
processes\cite{Wei:2009nc,Ke:2012wa,Ke:2013yka,Choi:2007se,Hwang:2006cua,Ke:2010vn,Ke:2011jf,Ke:2013zs,Ke:2011mu}.
The results obtained in this framework qualitatively agree with
the data for all the concerned processes.

For the transitions  $B_s\to \eta_c\phi$ and $B_s\to \eta_c
f_0(980)$ one needs to evaluate hadronic matrix elements $B_s\to
\phi$ and $B_s\to f_0(980)$.  The structure of
$f_0(980)$ is still not very clear yet, for example, Jaffe\cite{Jaffe:1976ig}
suggested $f_0(980)$ to be a four-quark state, instead, since the
resonance is close to the $K\bar{K}$ threshold a $K\bar{K}$
molecular structure was considered by Weinstein and
Isgur\cite{Weinstein:1982gc}. However, the regular $s\bar s$
structure for $f_0(980)$ still cannot be ruled out
\cite{Scadron:1982eg,Klabucar:2001gr,Cheng:2002ai}. In this paper the scalar meson
$f_0(980)$ is regarded as a conventional mixture of ${1\over\sqrt 2}(\bar uu+\bar dd)$ and $s\bar s$.

In Ref.\cite{Cheng:2003sm} the authors studied the formula of
$0^{-}\to 1^{-}$ and  $0^{-}\to 0^{+}$ in the LFQM. Actually, the
$0^{-}\to 1^{-}$ hadronic matrix element can be parameterized by
four form factors $A_0$, $A_1$, $A_2$ and $V$ and whereas for
$0^{-}\to 0^{+}$ transition it can be parameterized by two form
factors $F_0$ and $F_1$. Their detailed expressions obtained in
LFQM  can be found in Ref. \cite{Cheng:2003sm}.  In this work, we
will calculate these form factors numerically. With the form
factors one can further evaluate the transition widths of $B_s\to
\eta_c\phi$ and $B_s\to \eta_c f_0(980)$. In this model the
Gaussian-type wave functions are often used to depict the spatial
distribution of the inner constituents in the hadrons. There
exists a free parameter $\beta$ in the wave-function beside the
masses of the constituents. One should fix it by comparing the
decay constant of the involved meson which is either theoretically
calculated in LFQM with data.

This paper is organized as following: after this introduction, we
list all relevant formulas in Sec.II, and then in Sec.
III, we present our numerical results along with all inputs which
are needed for the numerical computations. In the last section we
draw our conclusion and make a brief discussion.

\section{The formulas for the decays of $B_s\to \eta_c\phi$ and $B_s\to \eta_c f_0(980)$ in LFQM}

\begin{center}
\begin{figure}[htb]
\begin{tabular}{c}
\scalebox{0.7}{\includegraphics{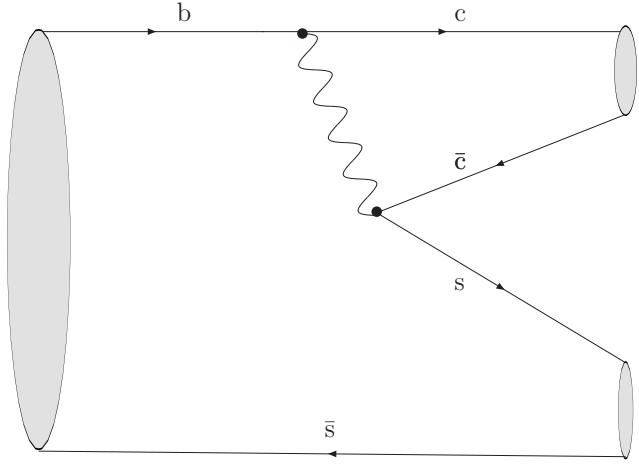}}
\end{tabular}
\caption{Feynman diagrams depicting the strong decay
.\label{fig1}}
\end{figure}
\end{center}

The leading contributions to $B_s\to \eta_c\phi$ and
$B_s\to \eta_c f_0(980)$  are shown in Fig.\ref{fig1}. We will
discuss them respectively in the following text.

\subsection{$B_s\rightarrow \phi$ transition in the LFQM}
The decay proceeds via $b\to \bar c c\bar s$ at tree level which
is an internal $W$-emission\cite{Aaij:2017hfc} process. The hadronic matrix element
is factorized as\cite{Bauer:1986bm}
\begin{eqnarray}\label{2s1}
\mathcal{A}=\frac{G_FV^*_{cs}V_{bc}\,a_2}{\sqrt{2}}\langle \eta_c
\phi|(\bar c c)_{V-A}(\bar s
b)_{V-A}|B_s\rangle=\frac{G_FV^*_{cs}V_{bc}\,a_2}{\sqrt{2}}\langle
\eta_c|(\bar c c)_{V-A}|0\rangle\langle \phi|(\bar s
b)_{V-A}|B_s\rangle,
\end{eqnarray}
where $a_2$ is the factor introduced in the introduction. It is also noted
the first term in Eq.(\ref{factor})
$<\eta_c\phi (f_0(980))| \bar c\gamma_{\mu}(1-\gamma_5)c|0><0|\bar s\gamma^{\mu}(1-\gamma_5)b|B_s>$ can be re-organized
via the crossing symmetry to a new form  which indeed
corresponds to a process where a $q\bar q$ pair annihilates into an $c\bar c$ pair. It is very suppressed, so we ignore this term
in later calculations.

The transition $B_s\rightarrow \phi$ is a typical process and
the involved form factors are defined as
 \be \label{eq1}
  \la V(P'',\vp^\pp)|V_\mu|P(P^\prime)\ra &=&
i\Bigg\{(M'+M'') \vp''^*_\mu A^{PV}_1(q^2)  - {\vp''^*\cdot
P'\over M'+M''}p_\mu A^{PV}_2(q^2) \non \\
&-& 2M'' {\vp''^*\cdot P'\over
q^2}q_\mu\left[A^{PV}_3(q^2)-A^{PV}_0(q^2)\right]\Bigg\},
\non \\
   \la V(P^\pp,\vp^\pp)|A_\mu|P(P^\prime)\ra &=& -{1\over
  M'+M''}\,\epsilon''_{\mu\nu\rho\sigma}\vp^{*\nu}\mathcal{P}^\rho
  q^{\sigma}V^{PV}(q^2),
 \en
with
 \be A^{PV}_3(q^2)=\,{M'+M''\over 2M''}\,A^{PV}_1(q^2)-{M'-M''\over
2M''}\,A^{PV}_2(q^2),
 \en
where $M' (M'')$  and $P' (P'')$ are the masses and momenta of the
vector  (pseudoscalar) states. We also set $\mathcal{P}=P'+P''$
and $q=P'-P''$.

In Ref.\cite{Cheng:2003sm} the authors deduce all
the expressions for the form factors $A_0$, $A_1$, $A_2$ and $V$
in the covariant LFQM. For example
\begin{eqnarray}\label{2s4}
V(q^2)=&&(M'+M'')\frac{N_c}{16\pi^3}\int dx_2d^2p'_\perp
 \frac{2h'_ph''_v}{x_2\hat{N}'_1\hat{N}''_1}
 \left\{x_2m_1'+x_1m_2+(m_1'-m_1'')\frac{p'_\perp\cdot
 q_\perp}{q^2}\right.\nonumber\\&&\left.
 +\frac{2}{w''_v}\left[p'^2_\perp+2\frac{(p'_\perp\cdot
 q_\perp)^2}{q^2}\right]\right\},\nonumber\\
\end{eqnarray}
where $m'_1,m''_1$ and $m_2$ are the corresponding quark masses,
$M'$ and $M''$ are the masses of the initial and final mesons
respectively. The wave functions are included in $h'_p$ and $h''_v$ and
they are usually chosen to be  Gaussian-type and the parameter $\beta$
in the Gaussian wave function is closely related to the confinement scale and
is expected to be of order $\Lambda_{\rm QCD}$. $N'_1$ and $N''_1$
come from the propagators of the inner quark or antiquark of the mesons. $N_c=3$
is the color factor. The notations $N'_1$, $N''_1$, $h'_p$ and $h''_v$
are given in the appendix.

One can refer to Eqs.(32) and (B4)
of Ref.\cite{Cheng:2003sm} for finding the explicit expressions of
$A_0$, $A_1$ and $A_2$ and the corresponding derivations.

\subsection{The transition $B_s\to f_0(980)$}\label{formula}

The amplitude for $B_s\to \eta_cf_0(980)$ is
\begin{eqnarray}\label{2s1}
\mathcal{A}=\frac{G_FV^*_{cs}V_{bc}\,a_2}{\sqrt{2}}\langle \eta_c
f_0(980)|(\bar c c)_{V-A}(\bar s
b)_{V-A}|B_s\rangle=\frac{G_FV^*_{cs}V_{bc}\,a_2}{\sqrt{2}}\langle
\eta_c|(\bar c c)_{V-A}|0\rangle\langle f_0(980)|(\bar s
b)_{V-A}|B_s\rangle.
\end{eqnarray}

$B_s\rightarrow f_0(980)$ is a typical $P\rightarrow S$ transition
process. The form factors for $P\rightarrow S$ are defined as
\begin{eqnarray}\label{2s1}
\langle
S(P'')|A_\mu|P(P')\rangle=i\left[u_+(q^2)p_\mu+u_-(q^2)q_\mu\right].
\end{eqnarray}

As an example, the explicit expression of $u_+$ is presented as
\begin{eqnarray}\label{2s4}
u_+(q^2)=&&\frac{N_c}{16\pi^3}\int dx_2d^2p'_\perp
 \frac{h'_ph''_s}{x_2\hat{N}'_1\hat{N}''_1}
 \left[-x_1(M'^2_0+M''^2_0)-x_2q^2+x_2(m_1'+m_1'')^2\right.\nonumber\\&&
 \left.+x_1(m'_1-m_2)^2+x_1(m''_1+m_2)^2\right],
\end{eqnarray}
where $h''_s$, $M'_0$ and $M''^2_0$ are given in the appendix.
The explicit expression of $u_{-}(q^2)$  is formulated in Ref.
\cite{Cheng:2003sm}.

As postulated, $f_0(980)$ is a pure $q\bar q$ state and its quark
structure is a superposition state as  $|f_0(980)>=\sin\theta
|{1\over\sqrt 2}(u\bar u+d\bar d)>+\cos\theta|s\bar s>$. Since
strange quark $s$ in $B_s$ can directly transit into the final
scalar meson as a spectator, one can notice that only $s\bar{s}$
component of $f_0(980)$ contributes to the transition $B_s \to
f_0(980)$. In
Ref.\cite{ElBennich:2008xy,Li:2008tk,Ghahramany:2009zz,Colangelo:2010bg}
the transition was studied using Covariant Light-Front Dynamics
(CLFD), Dispersion Relations (DR), PQCD approach, QCD sum rules
(QCDSR) and light-cone QCD sum rules (LCQCDSR). In those
articles\cite{ElBennich:2008xy,Li:2008tk,Ghahramany:2009zz,Colangelo:2010bg}
the form factors of the transition are defined as
\begin{eqnarray}\label{2s1}
\langle
S(P'')|A_\mu|P(P')\rangle=-i\left\{F_1(q^2)[\mathcal{P}_\mu-\frac{m_{B_s}^2-m^2_{f_0(980)}}{q^2}q_\mu]+F_0(q^2)\frac{m_{B_s}^2-m^2_{f_0(980)}}{q^2}q_\mu\right\}.
\end{eqnarray}
There are two relations $F_1=-u_+(q^2)$ and
$F_0=-[u_+(q^2)+\frac{q^2}{m_{B_s}^2-m^2_{f_0(980)}}u_-(q^2)]$ which associate the conventional form factors used in literature
with that we introduced above.

\subsection{Extension of the form factors to the physical region and the decay constant of $\eta_c$}
As discussed in Ref.\cite{Cheng:2003sm} the form factors are
calculated in the space-like region with $q^+=0$, thus to obtain
the physical amplitudes an extension to the time-like region is
needed. To make the extension one should write out an analytical
expressions for these form factors, and in Ref.\cite{Cheng:2003sm} a
three-parameter form was suggested
\begin{eqnarray}\label{eq2}
 F(q^2)=\frac{F(0)}{
  \left[1-a\left(\frac{q^2}{M_{B_s}^2}\right)
  +b\left(\frac{q^2}{M_{B_s}^2}\right)^2\right]}.
 \end{eqnarray}
where $F(q^2)$ denotes all $A_1{(q^2)}$, $A_2{(q^2)}$,
$A_3{(q^2)}$, $V{(q^2)}$, $F_1{(q^2)}$ and $F_0{(q^2)}$. $F(0)$ is
the value of $F(q^2)$ at $q^2=0$. In the scheme of LFQM one can
calculate $F(q^2)$ for the  space-like region ($q^2<0$), then through
Eq.(\ref{eq2}) $a$ and $b$ can be solved out. When we apply that
expression of $F(q^2)$ for $q^2>0$ with the same  $a$ and $b$, the
form factors are extrapolated to the time-like physical regions.
That is a natural analytical extension.

In the two processes, there is a unique matrix element $\langle
\eta_c|(\bar c c)_{V-A}|0\rangle$ which determines the decay
constant of $\eta_c$ and
\begin{eqnarray}
\langle\eta_c(p)|A_\mu|0\rangle=if_{\eta_c}p_\mu. \end{eqnarray}
Some  mesons' decay constants can be fixed by fitting data, whereas others must
be calculated in terms of phenomenological models or the lattice because no data are available so far.
Here the case for $f_{\eta_c}$  belongs to the latter.

In this scheme  $\langle \eta_c|(\bar c c)_{V-A}|0\rangle$ is
factorized out from the hadronic matrix element and is independent
of the matrix element $\langle f_0(980)|(\bar s
b)_{V-A}|B_s\rangle$. Moreover, if replacing $\langle \eta_c|(\bar c c)_{V-A}|0\rangle$
by $\langle J/\psi|(\bar c
c)_{V-A}|0\rangle$ which is related to the decay constant
$f_{J/\psi}$, one can study the transition $B_s\to J/\psi
f_0(980)$.

\section{Numerical results}
In this work, $m_s=0.37$ GeV, $m_c=1.4$ GeV and $m_b=4.64$ GeV are
adopted according to Ref. \cite{Cheng:2003sm}. $V_{cs}$ and
$V_{bc}$ are taken from the databook \cite{PDG12}. The parameter
$\beta$ in the wave function is fixed by calculating the
corresponding decay constant and comparing it with
data\cite{Cheng:2003sm}. For the vector meson $\phi$ one can
extract the decay constant ($227.7\pm1.2)$ MeV from the data
$BR(\phi\to e^+e^-)((2.954\pm0.030)\times 10^{-4})$\cite{PDG12}
and then $\beta_{\phi}=(0.3001\pm0.0010)$ GeV is achieved. For the
pesudoscalar meson $B_s$ its decay constant ($228.4\pm3.2$ MeV)
coming from the lattice result\cite{Aoki:2016frl} is used and we
obtain $\beta_{B_s}=(0.6165\pm0.0072)$ GeV.

In order to calculate the relevant form factors we need to know
$\beta^s_{f_0}$. For a scalar meson, as long as the masses of the
valence quark and antiquark are equal,  due to a symmetry with
respect to $x_1$ and $x_2$ which are their shares of  momenta in
the meson, the decay constant becomes zero as it should be. It is
shown by the integral over $x_1$ and $x_2$ in the framework of
LFQM \cite{Cheng:2003sm}. Following
Ref.\cite{Cheng:2003sm,Ke:2009ed}, we set $\beta^s_{f_0}=0.3$ in
our numerical computations. The mixing parameter $\theta$ takes a
value of $(56\pm6)^\circ$ which was fixed by fitting the branching
ratio of $D_s\to f_0(980)e^+\nu_e$\cite{Ke:2009ed} and then the
decay constant is $f_{\eta_c}=(387\pm7)$
MeV\cite{Becirevic:2013bsa}. It is also noted, when the
semileptonic decay of $D_s\to f_0(980)+e^+\nu_e$ was measured by
the CLEO collaboration, there were no data on $D_s\to
f_0(500)+e^+\nu_e$ available, therefore based on the mixing
postulation, such mixing angle was obtained by fitting only the
data of $B_s\to f_0(980)e^+\nu_e$. Later in this work, we will
show that the recent measurements on non-leptonic decays of
$B_s\to f_0(980)+X$ and $B_s\to f_0(500)+X$ disagree with the
mixing picture. We will give more discussions in the last section.

\begin{table}
\caption{the parameters $F_1(0),\ a ,\ b$ are defined in Eq.
(\ref{eq2}).} \label{table1}
\begin{ruledtabular}
 \begin{tabular}{ccccccccc}
 $F(q^2)$ &$F_1(0)$ &$a$& $b$\\\hline
 $A_0$ &0.292&1.590&1.794\\\hline
 $A_1$ &0.247&1.068&0.310 \\\hline
 $A_2$ &0.226&1.764&1.172 \\\hline
 $V$ &0.303&1.949&1.410 \\\hline
 $F_1$ &0.239{cos$\theta$}&1.690&0.917 \\\hline
 $F_0$ &0.239{cos$\theta$}&0.514&0.236 \\
\end{tabular}
\end{ruledtabular}
\end{table}
In Tab.\ref{table1} we present the parameters in those form
factors when all the input parameters are taking the central
values given elsewhere. In
Ref.\cite{ElBennich:2008xy,Li:2008tk,Ghahramany:2009zz,Colangelo:2010bg}
the transition $B_s\to f_0(980)$ were also studied and we  collect the
results  in Tab.\ref{table2}. Our prediction is close to the value
-0.238 obtained by the authors of \cite{Colangelo:2010bg} which includes the next-to-leading
order corrections.
\begin{table}
\caption{the $B_s\to f_0(980)$ form factor $F_0(q^2=0)$ with
$\cos\theta$=1.} \label{table2}
\begin{ruledtabular}
 \begin{tabular}{ccccccccc}
 &CLFD/DR& PQCD &QCDSR &LCQCDSR&this work\\\hline
$F_0$&0.40/0.29&0.35&0.12&0.238 &0.239 \\
\end{tabular}
\end{ruledtabular}
\end{table}

At first we explore whether using the value $a_2$ ($0.23\pm0.06$)
fixed in Ref.\cite{Cheng:1994fr} the predicted decay width can
meet the present data. With all the form factors and  parameters
as given above, we obtain the branching ratio
$BR(B_s\to\eta_c\phi)=(2.795\pm 1.652)\times 10^{-4}$ where the
errors come from the uncertainties of $\beta_{B_s},
\,\beta_{\phi},\,f_{\eta_c}$ and $a_2$, but mainly from $a_2$.
Apparently the estimate is smaller than the data $(5.01 \pm 0.53
\pm 0.27\pm0.63)\times 10^{-4}$, but as indicated above, the
theoretical errors are relatively large, so within a 2$\sigma$
tolerance, one still can count them as being consistent.  If we
deliberately vary the parameter $a_2$ within a reasonable range,
as setting $a_2=0.308\pm0.029$ the branching ratio
$BR(B_s\to\eta_c\phi)$ becomes $(5.012\pm0.863)\times 10^{-4}$
which is satisfactorily consistent with data.

Using the new value of $a_2$ let us evaluate the branching ratio
of $B_s\to\eta_c+f_0(980)$ and we obtain
$BR(B_s\to\eta_cf_0(980))=(1.591\pm0.568)\times 10^{-4}$. If one
applies this result to make a theoretical prediction on the
branching ratio of $B_s\to\eta_c\pi^+\pi^-$ by assuming the
$\pi^+\pi^-$ pair fully coming from an on-shell $f_0(980)$, he
will notice that the prediction is consistent with the present
measured value of $BR(B_s\to\eta_c\pi^+\pi^-)$. It seems that the
$\pi^+\pi^-$ pair in $B_s\to\eta_c\pi^+\pi^-$ mainly comes from
$f_0(980)$. But a discrepancy immediately emerges. In
Ref.\cite{Scadron:1982eg,Klabucar:2001gr,Cheng:2002ai} the authors
suggest that the scalar $f_0(500)(\sigma)$ is the complemental
state of $f_0(980)$. Thus the $s\bar s$ component of $f_0(500)$
which dominantly decays into $\pi\pi$ pairs, would play the same
role as that of $f_0(980)$. If simply setting $\theta=0$, we
calculate the branching ratio of $B_s\to\eta_c 0^+(s\bar s)$ ( i.e
$B_s\to\eta_c \pi^+\pi^-$ ) again. In that case we obtain
$BR(B_s\to\eta_c \pi^+\pi^-)=(5.089\pm1.022)\times 10^{-4}$ which
is about three larger than the data. This would raise a conflict
between theoretical prediction and experimental data.

Using the decay constant $f_{\psi}=416.3\pm 5.3$
MeV\cite{Colangelo:2010bg} we also estimate $BR(B_s\to J/\psi
f_0(980))=(1.727\pm0.615)\times 10^{-4}$ which is slightly larger
than the data $(1.19\pm0.22)\times 10^{-4}$\cite{PDG2016}, it seems OK, but
at the quark level, we have theoretically evaluate $BR(B_s\to J/\psi 0^+(s\bar
s))$ and gain it as $(5.523\pm1.103)\times 10^{-4}$ which leads $BR(B_s\to J/\psi
f_0(500))$ to be much larger than the upper limit $BR(B_s\to J/\psi
f_0(500))<1.7\times 10^{-6}$\cite{PDG2016}.

One possibility to pave the gap between theoretical prediction and
data is to assume an exotic structure for $f_0(980)$, namely is a
$K\bar K$ molecule state or tetraquark or  a mixture of them.
Using data of LHC whose integrated luminosity reaches  3 fb$^1$
the structure of  $\bar
{B}^0_s\to J/\psi\pi^+\pi^-$ was studied\cite{Aaij:2014emv} and
the mixing angle $\theta<7.7^\circ$(at 90$\%$ C.L.) which is
consistent with the prediction of the tetraquark
model\cite{Stone:2013eaa,Fleischer:2011au}. Apparently if the
upper-limit of the mixing angle is confirmed our prediction on
$BR(B_s\to \eta_c f_0(980))$ and $BR(B_s\to J/\psi f_0(980))$ will
be at least twice larger than the data so the $q\bar q$ structure
is disfavored.

\section{Summary}
In this work based on the postulation that $f_0(980)$ and
$f_0(500)$ are mixture of $\frac{1}{\sqrt{2}}(\bar uu+\bar dd)$
and $\bar ss$ we evaluate the decay widths of $B_s\to\eta_c\phi$
and $B_s\to\eta_cf_0(980)$ in LFQM. At the quark level the two
transitions proceed dominantly through  an internal $W-$ emission
sub-process $b\to\bar c c\bar s$. By the factorization assumption
the hadronic matrix element can factorized into a simple
transition matrix element multiplying by the decay constant of the
involved pseudoscalar meson. In this scenario the effective Wilson
coefficient factor $a_2$ plays a crucial role. By the naive
factorization $a_2$ is just related to $c_2+ c_1/3$ due to the
color rearrangement. However such naive combination is only a
rough approximation because some nonperturbative QCD effects would
get involved for a complete color rearrangement. The new
contribution is not universal for $B$ or $D$ decays. Thus
extracting the value of $a_2$ will provide us with  information
about the nonperturbative QCD effects in the corresponding decays
and even more.

In order to calculate the decay widths of $B_s\to\eta_c\phi$ and
$B_s\to\eta_cf_0(980)$ one needs to compute the transition hadronic matrix
elements $B_s\to\phi$ ($0^-\to 1^-$) and $B_s\to f_0(980)$ ($0^-\to
0^+$) which can be parametrized by several form factors. The
phenomenological model LFQM is employed to calculate these form
factors in this work. With the form factors and all the input parameters
we evaluate the rate of $B_s\to \eta_c\phi$ and obtain the value as
$\mathcal{BR}(B_s\to\eta_c\phi)=(2.795\pm1.652)\times 10^{-4}$
as $a_2$ taking value of $0.23\pm0.06$ as an input. If one admits that $a_2$ is a
free parameter, he can vary it to be $0.308\pm0.029$ and the obtained result is
compatible with the data.

Using the new $a_2$ we evaluate the branching ratio of $B_s\to
\eta_c f_0(980)$ with $\theta=56\pm 6^{\circ}$ and obtain it as
$BR(B_s\to\eta_cf_0(980))=(1.591\pm0.568)\times 10^{-4}$ which is
almost consistent with the present data. It seems the $\pi^+\pi^-$
only comes from $f_0(980)$. However, this assumption brings up
unacceptable consequence, that since $f_0(500)$ contains a large
fraction of $\bar s s$ (proportional to $\sin^2\theta$), the
contribution of $B_s\to \eta_cf_0(500)\to \eta_c\pi^+\pi^-$
becomes un-tolerably large as $(3.498\pm1.249)\times 10^{-4}$,
this number would lead to the branching ratio of $B_s\to
\eta_c\pi^+\pi^-$ to be roughly $5\times 10^{-4}$ which is roughly
3 times larger than the measured value.

Moreover, the recent measurements indicate the branching ratio of $B_s\to J/\psi+f_0(980)$ is $1.19\times 10^{-4}$
while $BR(B_s\to J/\psi+f_0(500)<1.7\times 10^{-6}$. The data imply that if the mixture scenario is correct, the mixing angle should
be smaller than $7.7^{\circ}$ instead of the large $56^{\circ}$. In other words $BR(B_s\to J/\psi+f_0(500)<1.7\times
10^{-6}$ implies that the fraction of $\bar ss$ in $f_0(500)$
should be very tiny.

If we accept the small mixing angle $\theta\sim 7.7^{\circ}$, we
obtain $BR(B_s\to\eta_cf_0(980)\to \eta_c\pi^+\pi^-)$ to be
$4.998\times 10^{-4}$, namely is consistent with the allegation
that the final $\pi^+\pi^-$ pair in $B_s\to\eta_c\pi^+\pi^- $ is
totally from $f_0(980)$, however, the theoretical picture is
surely disagreed by the data.

It is an obvious contradiction that in the mixing scenario, no
matter what value the mixing angle is adopted, the calculated
branching ratio for $B_s\to\eta_c\pi^+\pi^-$ is at least 3 times
larger than the data.

A synthesis of the measured branching ratio of
$BR(B_s\to\eta_c\pi^+\pi^-)\sim 1.76\times 10^{-4}$ and the data
$BR(B_s\to J/\psi f_0(500))<7.7^{\circ}$ determines no room for a
subprocess $B_s\to\eta_cf_0(500)\to \eta_c\pi^+\pi^-$. Namely if
the mixing scenario is adopted, no matter choosing what value for
the mixing angle, one cannot let the theoretically prediction meet
the data.

Therefore, under a complete consideration, one should draw a
conclusion that the main contents of $f_0(980)$ are not a mixture
of $(\bar uu+\bar dd)/\sqrt 2$ and $\bar s s$, but could be a four
quark state: $K\bar K$ molecule as Isgur et al. suggested or a
tetraquark.

We suggest the experimentalists to carry out a more precise
measurement on the $B_s\to\eta_c\pi^+\pi^-$ where the invariant mass
of $\pi^+\pi^-$ would clearly tell us if $\pi^+\pi^-$ mainly
come from $f_0(980)$.

\section*{Acknowledgement}

This work is supported by the National Natural Science Foundation
of China (NNSFC) under the contract No. 11375128 and 11675082.

\appendix

\appendix

\section{Notations}

Here we list some variables appearing in the context.  The
incoming  meson in Fig. \ref{fig1} has the momentum $P'=p'_1+p'_2$
where $p'_1$ and $p'_2$ are the momenta of the off-shell quark and
antiquark and
\begin{eqnarray}\label{app1}
&& {p'}_1^+=x_1{P'}^+, \qquad ~~~~~~{p'}_2^+=x_2P^+, \nonumber\\
&& {p'}_{1\perp}=x_1{P'}_{\perp}+p'_\perp, \qquad
 {p'}_{2\perp}=x_2{P'}_{\perp}-p'_\perp,
 \end{eqnarray}
with  $x_i$ and $p'_\perp$ are internal variables and $x_1+x_2=1$.

The variables ${M'}_0$, $\tilde {{M'}_0}$, $h_p'$, $h_s'$,
$\hat{N_1'}$ and $\hat{N_1''}$ are defined as
\begin{eqnarray}\label{app2}
&&{M'}_0^2=\frac{{p'}^2_\perp+{m'}^2_1}{x_1}+\frac{{p'}^2_\perp+{m'}^2_2}{x_2},\nonumber\\&&
\tilde
{{M'}_0}=\sqrt{{M'}_0^2-({m'}_1-{m'}_2)^2},\nonumber\\&&h_p'=(M'^2-M'^2_0)\sqrt{\frac{x_1x_2}{N_c}}\frac{1}{\sqrt{2}\tilde{M_0'}}\varphi',\nonumber\\&&
h_s'=(M'^2-M'^2_0)\sqrt{\frac{x_1x_2}{N_c}}\frac{\tilde{M_0'}}{2\sqrt{6}M_0'}\varphi'_p,\nonumber\\&&\hat{N_1'}=x_1(M'^2-M'^2_0),\nonumber\\&&
\hat{N_1''}=x_1(M''^2-M''^2_0).
 \end{eqnarray}
where
\begin{eqnarray}\label{app4}
\varphi'=4(\frac{\pi}{\beta^2})^{3/4}\sqrt{\frac{d{p'}_z}{dx_2}}{\rm
 exp}(-\frac{{p'}^2_z+{p'}^2_\perp}{2\beta^2}),\,\varphi'_p=\sqrt{2/\beta}\varphi',
 \end{eqnarray}
with
${p'}_z=\frac{x_2{M'}_0}{2}-\frac{{m'}_2^2+{p'}^2_\perp}{2x_2{M'}_0}$.



\begin{thebibliography}{99}

\bibitem{Aaij:2017hfc}
  R.~Aaij {\it et al.} [LHCb Collaboration],
  arXiv:1702.08048 [hep-ex].  

\bibitem{Isgur:1989vq}
  N.~Isgur and M.~B.~Wise,
  Phys.\ Lett.\ B {\bf 232}, 113 (1989).
  doi:10.1016/0370-2693(89)90566-2

\bibitem{Georgi:1990um}
  H.~Georgi,
  Phys.\ Lett.\ B {\bf 240}, 447 (1990).
  doi:10.1016/0370-2693(90)91128-X


\bibitem{Bauer:1986bm}
  M.~Bauer, B.~Stech and M.~Wirbel,
   Z.\ Phys.\ C {\bf 34}, 103 (1987).  doi:10.1007/BF01561122

\bibitem{Cheng:1986an}
  H.~Y.~Cheng,
  Z.\ Phys.\ C {\bf 32}, 237 (1986).
  doi:10.1007/BF01552501


\bibitem{Li:1988hr}
  X.~q.~Li, T.~Huang and Z.~c.~Zhang,
  Z.\ Phys.\ C {\bf 42}, 99 (1989).
  doi:10.1007/BF01565132


\bibitem{Cheng:1994fr}
  H.~Y.~Cheng and B.~Tseng,
   Phys.\ Rev.\ D {\bf 51}, 6259 (1995)  doi:10.1103/PhysRevD.51.6259  [hep-ph/9409408].



\bibitem{Jaus:1999zv}
  W.~Jaus,
  Phys.\ Rev.\  D {\bf 60}, 054026 (1999).
\bibitem{Cheng:2003sm}
  H.~Y.~Cheng, C.~K.~Chua and C.~W.~Hwang,
  Phys.\ Rev.\  D {\bf 69}, 074025 (2004).


\bibitem{Wei:2009nc}
  Z.~T.~Wei, H.~W.~Ke and X.~F.~Yang,
  Phys.\ Rev.\  D {\bf 80}, 015022 (2009)
  [arXiv:0905.3069 [hep-ph]].

\bibitem{Ke:2013yka}
  H.~W.~Ke, T.~Liu and X.~Q.~Li,
  Phys.\ Rev.\ D {\bf 89}, 017501 (2014)  [arXiv:1307.5925 [hep-ph]].  


\bibitem{Ke:2012wa}
  H.~W.~Ke, X.~H.~Yuan, X.~Q.~Li, Z.~T.~Wei and Y.~X.~Zhang,
   Phys.\ Rev.\ D {\bf 86}, 114005 (2012)  [arXiv:1207.3477 [hep-ph]].  





\bibitem{Choi:2007se}
  H.~M.~Choi,
  Phys.\ Rev.\  D {\bf 75}, 073016 (2007)
  [arXiv:hep-ph/0701263];

  H.~M.~Choi,
  J.\ Korean Phys.\ Soc.\  {\bf 53}, 1205 (2008)
  [arXiv:0710.0714 [hep-ph]].

\bibitem{Hwang:2006cua}
  C.~W.~Hwang and Z.~T.~Wei,
   J.\ Phys.\ G {\bf 34} (2007) 687  [hep-ph/0609036].  

\bibitem{Ke:2010vn}
  H.~W.~Ke, X.~Q.~Li, Z.~T.~Wei and X.~Liu,
  Phys.\ Rev.\  D {\bf 82}, 034023 (2010)
  [arXiv:1006.1091 [hep-ph]].

\bibitem{Ke:2011jf}
  H.~W.~Ke and X.~Q.~Li,
  Phys.\ Rev.\ D {\bf 84}, 114026 (2011)  [arXiv:1107.0443 [hep-ph]].  

\bibitem{Ke:2013zs}
  H.~W.~Ke, X.~Q.~Li and Y.~L.~Shi,
  Phys.\ Rev.\ D {\bf 87}, 054022 (2013)  [arXiv:1301.4014 [hep-ph]].  



\bibitem{Ke:2011mu}
  H.~W.~Ke and X.~Q.~Li,
   Eur.\ Phys.\ J.\ C {\bf 71}, 1776 (2011)  [arXiv:1104.3996 [hep-ph]].  

\bibitem{Jaffe:1976ig}
  R.~L.~Jaffe,
  Phys.\ Rev.\ D {\bf 15}, 267 (1977).
  doi:10.1103/PhysRevD.15.267

\bibitem{Weinstein:1982gc}
  J.~D.~Weinstein and N.~Isgur,
  Phys.\ Rev.\ Lett.\  {\bf 48}, 659 (1982).
  doi:10.1103/PhysRevLett.48.659



\bibitem{Scadron:1982eg}
  M.~D.~Scadron,
  Phys.\ Rev.\ D {\bf 26}, 239 (1982).
  doi:10.1103/PhysRevD.26.239

\bibitem{Klabucar:2001gr}
  D.~Klabucar, D.~Kekez and M.~D.~Scadron,
  J.\ Phys.\ G {\bf 27}, 1775 (2001)
  doi:10.1088/0954-3899/27/8/307
  [hep-ph/0101324].

\bibitem{Cheng:2002ai}
  H.~Y.~Cheng,
  Phys.\ Rev.\ D {\bf 67}, 034024 (2003)
  doi:10.1103/PhysRevD.67.034024
  [hep-ph/0212117].

\bibitem{ElBennich:2008xy}
  B.~El-Bennich, O.~Leitner, J.-P.~Dedonder and B.~Loiseau,
  Phys.\ Rev.\ D {\bf 79}, 076004 (2009)
  doi:10.1103/PhysRevD.79.076004
  [arXiv:0810.5771 [hep-ph]].

\bibitem{Li:2008tk}
  R.~H.~Li, C.~D.~Lu, W.~Wang and X.~X.~Wang,
  Phys.\ Rev.\ D {\bf 79}, 014013 (2009)
  doi:10.1103/PhysRevD.79.014013
  [arXiv:0811.2648 [hep-ph]].

\bibitem{Ghahramany:2009zz}
  N.~Ghahramany and R.~Khosravi,
  Phys.\ Rev.\ D {\bf 80}, 016009 (2009).
  doi:10.1103/PhysRevD.80.016009

\bibitem{Colangelo:2010bg}
  P.~Colangelo, F.~De Fazio and W.~Wang,
  Phys.\ Rev.\ D {\bf 81}, 074001 (2010)
  doi:10.1103/PhysRevD.81.074001
  [arXiv:1002.2880 [hep-ph]].




\bibitem{PDG12}
  K.~A.~Olive {\it et al.}  [Particle Data Group Collaboration],
  Chin.\ Phys.\ C {\bf 38}, 090001 (2014).  

\bibitem{Aoki:2016frl}
  S.~Aoki {\it et al.},
   Eur.\ Phys.\ J.\ C {\bf 77}, no. 2, 112 (2017)  doi:10.1140/epjc/s10052-016-4509-7  [arXiv:1607.00299 [hep-lat]].

\bibitem{Ke:2009ed}
  H.~W.~Ke, X.~Q.~Li and Z.~T.~Wei,
  Phys.\ Rev.\ D {\bf 80}, 074030 (2009)
  doi:10.1103/PhysRevD.80.074030
  [arXiv:0907.5465 [hep-ph]].



\bibitem{Becirevic:2013bsa}
  D.~Becirevic, G.~Duplancic, B.~Klajn, B.~Melic and F.~Sanfilippo,
  Nucl.\ Phys.\ B {\bf 883}, 306 (2014)
  doi:10.1016/j.nuclphysb.2014.03.024
  [arXiv:1312.2858 [hep-ph]].

\bibitem{PDG2016}
  C.~Patrignani {\it et al.} [Particle Data Group],
  Chin.\ Phys.\ C {\bf 40}, no. 10, 100001 (2016).
  doi:10.1088/1674-1137/40/10/100001

\bibitem{Aaij:2014emv}
  R.~Aaij {\it et al.} [LHCb Collaboration],
  Phys.\ Rev.\ D {\bf 89}, no. 9, 092006 (2014)
  doi:10.1103/PhysRevD.89.092006
  [arXiv:1402.6248 [hep-ex]].

\bibitem{Stone:2013eaa}
  S.~Stone and L.~Zhang,
  Phys.\ Rev.\ Lett.\  {\bf 111}, no. 6, 062001 (2013)
  doi:10.1103/PhysRevLett.111.062001
  [arXiv:1305.6554 [hep-ex]].

\bibitem{Fleischer:2011au}
  R.~Fleischer, R.~Knegjens and G.~Ricciardi,
  Eur.\ Phys.\ J.\ C {\bf 71}, 1832 (2011)
  doi:10.1140/epjc/s10052-011-1832-x
  [arXiv:1109.1112 [hep-ph]].

\end{thebibliography}
\end{document}